\documentclass{gistt}

\usepackage{graphicx} 
\usepackage{paralist}

\usepackage{listings}
\usepackage{tikz}
\usetikzlibrary{shapes,arrows}
\usetikzlibrary{calc}
\usetikzlibrary{fit,positioning}
\usetikzlibrary{patterns}

\bibliography{references.bib}

\title{Instrumentation of Software Systems with OpenTelemetry\\ for Software Visualization}
\author{Malte Hansen\\
Kiel University, Kiel, Germany
\and
Wilhelm Hasselbring\\
Kiel University, Kiel, Germany}

\begin{document}

\maketitle

\begin{abstract}
As software systems grow in complexity, data and tools that provide valuable insights for easier program comprehension become increasingly important.
OpenTelemetry has become a standard for the collection of monitoring data.
In this work we present our experiences with different ways how OpenTelemetry can be leveraged to automatically instrument software systems for the purpose of software visualization.
Particularly, we explore automatic instrumentation with the OpenTelemetry SDKs, and both application and unit test instrumentation with the Java agent \mbox{inspectIT Ocelot}.
The collected data is exported to our live trace visualization tool ExplorViz.

\end{abstract}

\section{Introduction}
Monitoring data can be used to gain valuable insight into the behavior of software systems and thereby facilitate program comprehension.
OpenTelemetry is a collection of APIs, SDKs, and tools for software observability, and the de facto standard for enabling interoperability between tools that collect or process monitoring data~\cite{janes2023}.

In this paper we report on different ways how OpenTelemetry can be leveraged to collect tracing data.
In particular, we take a look at OpenTelemetry's own automatic instrumentation and both application and unit test monitoring with the inspectIT Ocelot\footnote{\url{https://www.inspectit.rocks/}} Java agent.
The instrumentation process and the resulting data are put into perspective from the point of view of our software visualization tool ExplorViz.

The rest of this paper is organized as follows:
In the next section we describe our software visualization approach.
We then look at three approaches for the collection of OpenTelemetry-compliant monitoring data in the context of software visualization.
Afterwards, we present previous publications which are related to our presented approaches.
We conclude the paper with a summary and brief outlook.

\section{Software Visualization Approach}
\begin{figure*}
	\includegraphics[width=\textwidth]{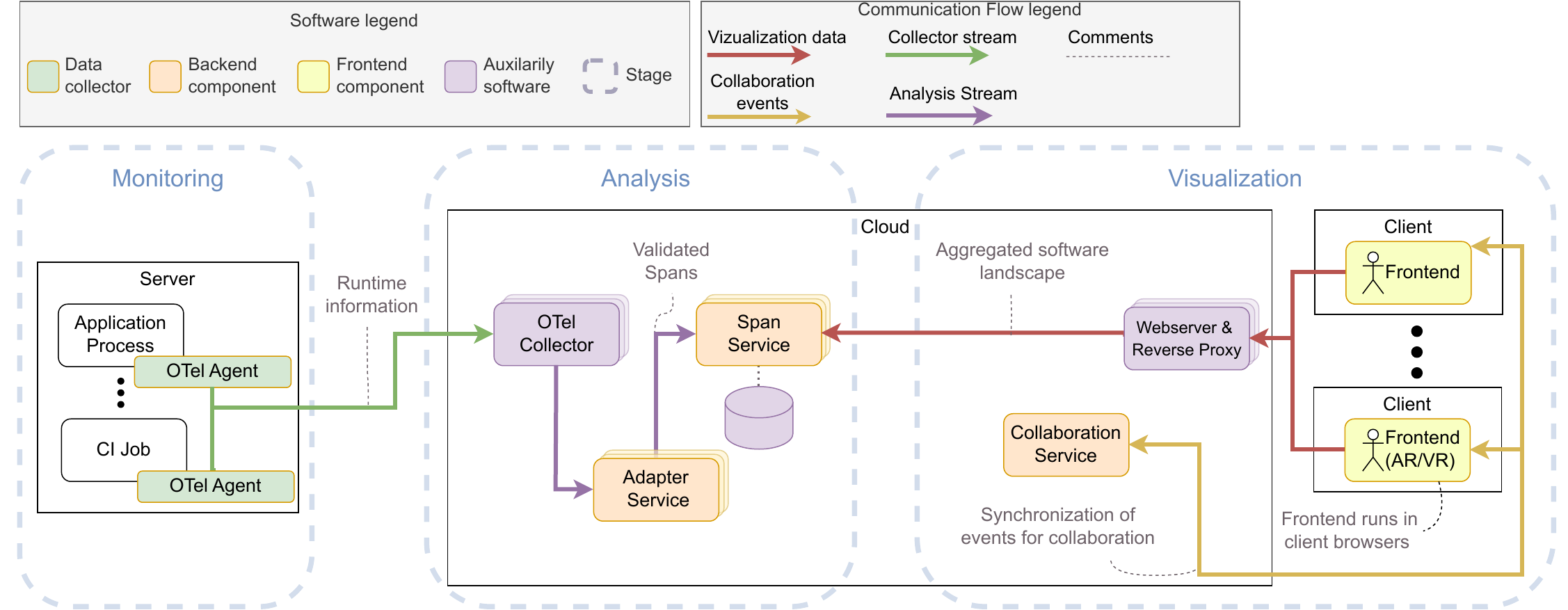}
	\caption{Simplified architecture diagram of ExplorViz depicting data flows.}
	\label{fig:workflow}
\end{figure*}

ExplorViz~\cite{hasselbring2020explorviz} is our web-based software visualization tool which employs the city metaphor~\cite{wettel2007}.
Up until a few years ago, Kieker~\cite{hasselbring2020kieker} was used as the sole monitoring tool for ExplorViz.
The compatibility between the two tools was lost when ExplorViz switched to the emerging and  popular OpenTelemetry~\cite{Blanco2023} standard for tracing as part of a major restructuring of ExplorViz.
In Figure \ref{fig:workflow} the conceptual architecture of ExplorViz and the flow of monitoring data is depicted.
Running applications are instrumented with an OpenTelemetry compliant agent in the monitoring stage.
The data is then exported via gRPC to an OpenTelemetry Collector which exports it to the backend of ExplorViz via Apache Kafka.\footnote{\url{https://kafka.apache.org/}}
The retrieved tracing data is processed in the analysis stage, where traces and structural data of the software system is reconstructed from the retrieved OpenTelemetry spans.
At last, both the processed dynamic and static data is requested and visualized by the frontend.

The resulting visualization by the example of a distributed version of the Spring PetClinic\footnote{\url{https://github.com/spring-petclinic/spring-petclinic-microservices}} is presented in Figure \ref{fig:explorviz}.
Applications are depicted by gray foundations.
The hierarchical package structure is depicted by green and blue components which can be interpreted as districts of the software city.
Components can contain classes, which are represented by blue rectangular cuboids and can be interpreted as buildings.
The communication, i.e. method calls, which occured in the observed time frame are depicted with yellow arcs.

The visualization can be explored collaboratively and supports diverse devices including augmented and virtual reality headsets via WebXR.

\section{Monitoring Approaches}
In this section we present different ways to monitor software systems for the purpose of software visualization.
First, we take a look at the automatic instrumentation solution provided by OpenTelemetry before we present application and unit test monitoring with the third party Java agent inspectIT Ocelot.

\subsection{Distributed Tracing}
Aside from the option to manually instrument an application, OpenTelemetry provides eleven language SDKs to automatically collect traces of software systems, whereby the SDK for Rust is in a beta status at the time of writing.
It is noteworthy that the automatic instrumentation of those SDKs is used solely for distributed tracing.
Therefore, function calls of software libraries and frameworks that handle distributed events such as HTTP requests or I/O-operations may be caught.

For ExplorViz, we explored the reference implementation for the instrumentation of JavaScript applications by using mongo-express\footnote{\url{https://github.com/mongo-express/mongo-express}} as a Node.js\footnote{\url{https://nodejs.org}}, and the Angular frontend for the Spring PetClinic~\footnote{\url{https://github.com/spring-petclinic/spring-petclinic-angular}} as a web-based JavaScript application as examples.
As implied by the distributed focus of the SDKs, function calls that occur solely within an application, often including most parts of the business logic, are usually not captured.
As there is also no information about the source code file that the function call originates from, cluster algorithms would be needed to create an artificial software structure.
On its own, the solely distributed instrumentation is not sufficient for a comprehensive visualization with ExplorViz, since our visualization approach is focused around the internal behavior and structure by means of the software city metaphor.

On the other hand, internal function calls can be monitored with Kieker or inspectIT Ocelot, which is considered in the two upcoming sections.
\begin{figure}[htbp]
	\includegraphics[width=\textwidth/2]{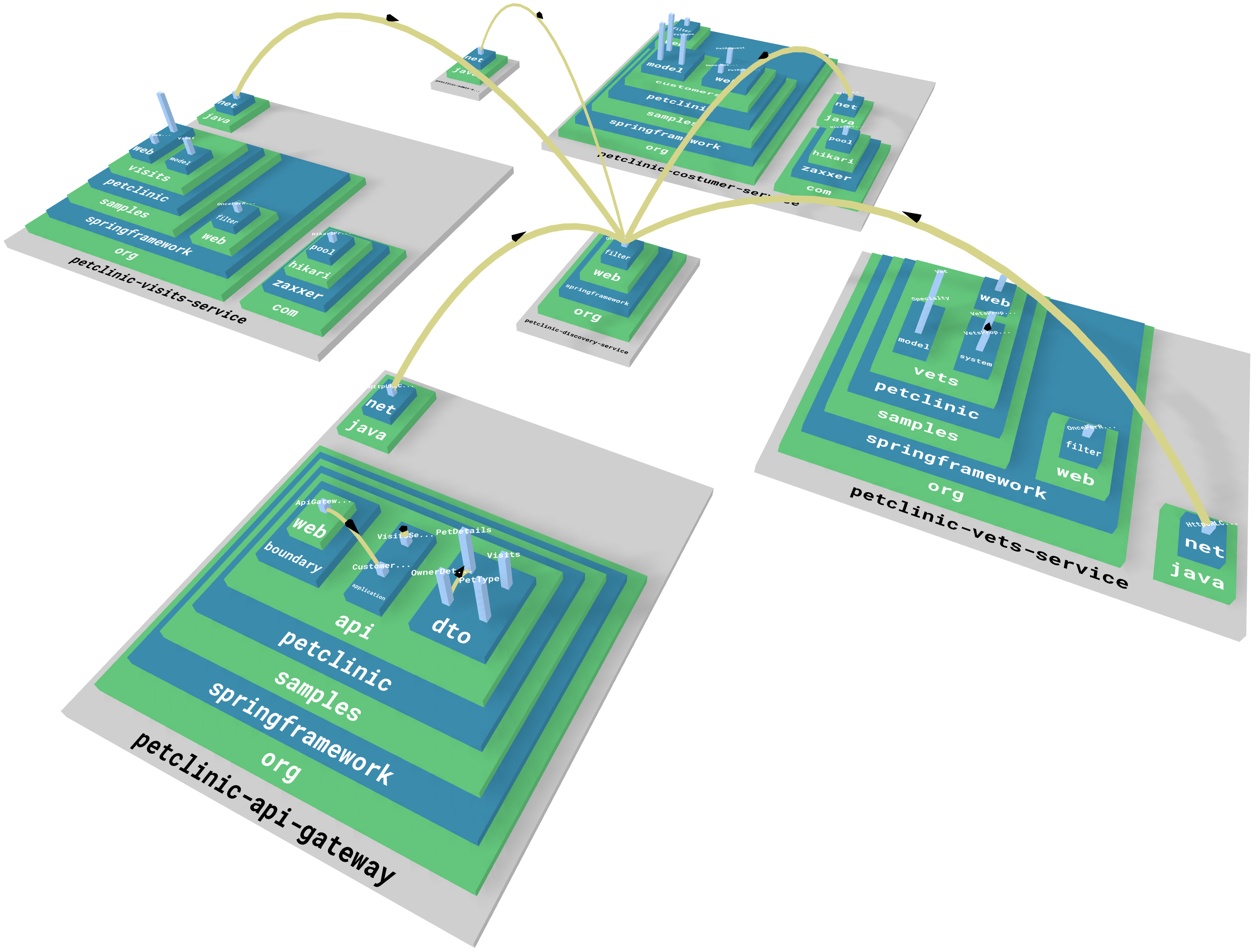}
	\caption{Distributed version of the Spring PetClinic as visualized by ExplorViz with data from the \mbox{inspectIT} Ocelot agent.}
	\label{fig:explorviz}
\end{figure}

\subsection{Application Monitoring}
As ExplorViz is built upon the city metaphor, collecting monitoring data about the inner structure and behavior of a software system is the key to a comprehensive visualization.
The Java agent inspectIT Ocelot can dynamically instrument Java applications and export the collected monitoring data in accordance with the OpenTelemetry standard.
The agent can be started with a given software or added to a running process of the java virtual machine.
For easy deployment, all components of ExplorViz are developed with continuous integration and continuous delivery as Docker containers in mind.
Therefore, to instrument a given software, the inspectIT Ocelot agent along with a configuration for instrumentation is mounted into the software's container.
In the end, only the container's entrypoint needs to be adapted to start with the agent, leaving the monitored software untouched and resulting in an automatic instrumentation process.

An example of a software landscape visualized using this monitoring approach is presented in Figure \ref{fig:explorviz}.
As can be seen, this approach has the advantage that both internal and distributed communication is captured. 
Therefore, this is the current default monitoring approach for ExplorViz.

\subsection{Monitoring of Unit Tests}
As a third option for monitoring software systems, we propose the instrumentation of unit tests.
The setup of the previous section can be adapted such that inspectIT Ocelot is started with the test suite, for example, as part of a Gradle project.
This could be used as a supplement to static code analysis as part of continuous integration to detect unexpected changes in program behavior beyond the binary results of unit tests.
However, this approach comes with some caveats.
From a technical perspective, the Java agent needs to be fully loaded before tests are executed to avoid that test cases are missed.
The execution of all unit tests can then result in millions of spans in a short amount of time for large projects like Spring for Apache Kafka.\footnote{\url{https://github.com/spring-projects/spring-kafka}}
Therefore, it can easily happen that spans are dropped, for example at the OpenTelemetry Collector, which acts as a buffer.
More fundamentally, this approach relies on the expressiveness and comparability of unit tests, which can vary greatly from project to project~\cite{reichelt2023}.

\section{Related Work}
Kieker~\cite{Kieker2012,hasselbring2020kieker} is a tool that provides dynamic analysis of software systems.
Just as inspectIT Ocelot, it provides data about the function calls that occur both inside and between applications.
As opposed to only being able to instrument Java applications, Kieker supports several programming languages such as C++ and has included components to analyze the gathered monitoring data.
On the other hand, as of now, Kieker uses its own data format for monitoring data and thus is not compatible with tools that expect OpenTelemetry spans, such as ExplorViz.
However, it is an ongoing effort to restore the compatibility between Kieker and ExplorViz by adding an OpenTelemetry compliant exporter to Kieker.

Reichelt et al.~\cite{reichelt2021overhead} compared the overhead of OpenTelemetry, inspectIT, and Kieker.
They observed that Kieker is creating slightly less overhead than inspectIT and OpenTelemetry
when processing traces.
Reichelt~\cite{reichelt2023} also examined performance changes on source code level.
Just as we did, he also instrumented unit tests to gain insights into the software system.
However, he used Kieker as opposed to inspectIT Ocelot and had a greater focus on regression testing for performance than the behavior-driven software visualization approach that we pursue.

\section{Summary and Outlook}
We presented three approaches to gather tracing data from software systems in accordance with the OpenTelemetry standard.
We looked at the resulting data from the point of view of our software visualization tool ExplorViz, which is centered around the metaphor of software cities.
Both the simple to setup automatic instrumentation of OpenTelemetry and the instrumentation of unit tests have distinct use cases.
However, our experience has shown that a monitoring tool which can automatically track both distributed and internal traces is best suited for our visualization approach.
As of now, the Java agent inspectIT Ocelot is the best fit for ExplorViz.

We plan to extend the support for OpenTelemetry of ExplorViz to also include metrics and deployment information such as data about Kubernetes clusters.
At last, we are looking forward to regain full interoperability between ExplorViz and Kieker, to leverage its powerful monitoring capabilities and eventually use it in place of inspectIT Ocelot.

\section{Acknowledgements}
We want to thank Roman Hemens and Piet Schumacher for their contributions to ExplorViz regarding OpenTelemetry.

\printbibliography

\end{document}